\documentclass[aps,twocolumn,prl,reprint,preprintnumbers,superscriptaddress,nofootinbib]{revtex4-2}
\usepackage{amsmath,amssymb,amsthm,etoolbox,tikz,tikz-cd,multirow,bm,graphicx,array,pgf,soul,bm,xcolor,comment,footnote}
\pdfoutput=1
\usepackage[all]{xy}
\usepackage[shortlabels]{enumitem}
\usepackage[normalem]{ulem}
\usepackage{booktabs}
\usepackage[hidelinks]{hyperref}
\theoremstyle{plain}
\newtheorem*{thm*}{Theorem}

\newtheorem*{proposal*}{Proposal}
\begin{document}
\title{
Algebro-geometric bootstrapping
from OPE decoupling
}

\vspace*{-3cm} 
\begin{flushright}
{\tt CAL-TH-2026-006}\\
{\tt DESY-26-015}\\
\end{flushright}
\author{Monica Jinwoo Kang}
\email[\texttt{monicak@tamu.edu}]{}
\affiliation{Mitchell Institute for Fundamental Physics and Astronomy, Texas A\&M University, College Station, TX 77843, U.S.A.}

\author{Craig Lawrie}
\email[\texttt{craig.lawrie1729@gmail.com}]{}
\affiliation{Deutsches Elektronen-Synchrotron DESY, Notkestrasse 85, 22607 Hamburg, Germany}

\author{Jaewon Song}
\email[\texttt{jaewon.song@kaist.ac.kr}]{}
\affiliation{Department of Physics, Korea Advanced Institute of Science and Technology, Daejeon 34141, Republic of Korea}
\affiliation{Walter Burke Institute for Theoretical Physics, California Institute of Technology, Pasadena, CA 91125, U.S.A.}

\begin{abstract}
\noindent 
We conjecture that decoupling relations in the operator product expansion of a 4d $\mathcal{N}=2$ superconformal field theory (SCFT) are encoded by an algebro-geometric object: a bifiltered affine scheme. We demonstrate how this scheme reproduces the Macdonald index (thus the Schur index) as well as the Higgs branch. Although the associated scheme typically admits continuous deformations, we find that a geometric extremization principle uniquely fixes these moduli, thereby providing a possible geometric route toward a classification of 4d $\mathcal{N}=2$ SCFTs.
\end{abstract}

\maketitle






\section{Introduction}

A central theme in modern quantum field theory (QFT) is that many consistent QFTs exist beyond the reach of conventional Lagrangian descriptions \cite{Argyres:1995jj, Gaiotto:2009we} and defy perturbative understanding. Therefore, it is desirable to search for a more robust principle that specifies a QFT. One such principle is the decoupling or truncation of operator products, which dramatically restricts the possible field theory data. This mechanism appears in various settings such as 2d minimal models \cite{Belavin:1984vu}, the 3d Ising model \cite{El-Showk:2014dwa, Chang:2024whx}, and 4d $\mathcal{N}=2$ SCFTs \cite{Liendo:2015ofa, Agarwal:2018zqi, Song:2021dhu}. This phenomenon does not require a Lagrangian description and is inherently non-perturbative. 

In general, a local conformal field theory is determined by the scaling dimensions and three-point OPE coefficients of all conformal primary operators, known as the conformal data. It is believed that this is a highly redundant description of a theory.
For example, the conformal bootstrap program has provided evidence that a 3d CFT with a $\mathbb{Z}_2$ global symmetry and precisely two relevant scalar operators, of opposite $\mathbb{Z}_2$-charge, is necessarily the 3d Ising model. The $\mathbb{Z}_2$ implies an OPE decoupling: the $\mathbb{Z}_2$-odd scalar decouples from the OPE of two $\mathbb{Z}_2$-even scalars. Furthermore, the extremal-functional method \cite{El-Showk:2012vjm} can be used to recover the conformal data.

Based on these observations, we raise the following question: can we employ this decoupling phenomenon to define a local quantum field theory more generally? In our recent work \cite{Kang:2025zub}, we answered this question in the affirmative, at least for determining a protected subsector of a 4d $\mathcal{N}=2$ SCFT. The most transparent example is the $(A_1, A_{2n})$ Argyres--Douglas theory \cite{Argyres:1995jj}, where OPE decoupling occurs in the $(n+1)$-th product of the stress tensor \cite{Liendo:2015ofa, Agarwal:2018zqi}, leading to an algebro-geometric object:
\begin{equation}
	T^{n+1} \sim 0 \quad \leadsto \quad R = \mathbb{C}[x]/(x^{n+1}) \,.
\end{equation}
$X=\operatorname{Spec} R$ is a \emph{fat point}, which is trivial as a variety but non-trivial as a \emph{scheme}, retaining the nilpotent structure dictated by the OPE decoupling. The Macdonald index, which counts quarter-BPS operators \cite{Gadde:2011ik, Gadde:2011uv}, can be obtained by computing the Hilbert series of an associated geometric object, the arc space $J_\infty X$, providing an algebro-geometric interpretation of indices \cite{Andrews:2025krn,Kang:2025zub,Bhargava:2023hsc}.

Following the logic that OPE decoupling in 4d $\mathcal{N}=2$ SCFTs is captured by an algebraic scheme, it is natural to reverse the direction and ask: what conditions must a scheme satisfy such that it captures OPE decoupling? Understanding such conditions would open a path toward defining and classifying (generally non-Lagrangian) 4d $\mathcal{N}=2$ SCFTs in purely algebro-geometric terms. 

We find that a crucial condition follows from an extremization principle, which we determine for $(A_{k-1}, A_{N-1})$ Argyres--Douglas theories \cite{Argyres:1995jj, Argyres:1995xn, Xie:2012hs}. While a given bifiltered affine scheme typically admits continuous deformations, we find that generic points in this moduli space lead to spectra that we cannot interpret as those of a 4d SCFT. Remarkably, only at unique extremal points, characterized by a change in the topology of the associated jet scheme, does the spectrum coincide with that of a known physical theory. We interpret this extremality as a physical consistency condition that selects viable SCFTs from a larger space of geometric data.

Concretely, we propose the following.
\begin{proposal*}
    For an ``extremal'' bifiltered affine scheme $X$, with coordinate ring $R=\mathbb{C}[X]$, we can associate a 4d $\mathcal{N}=2$ SCFT, whose Macdonald index is
    \begin{equation}\label{eqn:prop}
        I_\text{Mac}(q, T) = \operatorname{HS}_{q,q,T}\left(\operatorname{gr}(J_\infty(R))\right) \,,
    \end{equation}
    and the Higgs branch is 
    \begin{equation}\label{eqn:prop2}
        \mathcal{M}_H = (X)_{\text{red}} = \operatorname{Spec}(R)_\text{red} \,.
    \end{equation}
\end{proposal*}
\noindent The choice of geometry stems from the OPE decoupling structure, and the bifiltration structure is motivated by the superconformal symmetry. The extremality condition provides a way to pinpoint a physically realizable theory. We will provide precise definitions of the terms used above in the following sections. 

\section{Arc spaces and Hilbert series}\label{sec:prelim}

In this section, we introduce the mathematical preliminaries necessary for our proposal that a bifiltered ring, the coordinate ring of a bifiltered affine scheme, serves as both the scheme-theoretic enhancement of the Higgs branch moduli space and as the generating object for the Macdonald index. We typically denote the bifiltered ring simply by $R$, leaving the bifiltration structure implicit.

First, a bifiltered ring is a ring $R$ together with a sequence of subgroups
\begin{equation}
    F^{p,q}\,R \subseteq R \quad \text{for} \quad p,q \in \mathbb{Z} \,.
\end{equation}
These subgroups must satisfy three conditions:
\begin{enumerate}
    \item $\displaystyle
        F^{p,q}\,R \subseteq F^{p',q'}\,R \quad \text{ if } p \leq p', q \leq q'$,
    \item $\displaystyle \bigcup_{p,q} F^{p,q}\,R = R $,
    \item $\displaystyle F^{p,q}\,R\, \cdot F^{p',q'}\,R \subseteq F^{p+p',q+q'}\,R$.
\end{enumerate}
An $\ell$-filtered ring for $\ell \geq 1$ integer satisfies the obvious generalizations of these conditions.

There is a simple way to specify a bifiltration. Consider the bigraded polynomial ring
\begin{equation}
    S = \mathbb{C}[x_1, \cdots, x_m] = \bigoplus_{p,q}S_{p,q} \,,
\end{equation}
where the bigrading is specified by the bidegrees: $\operatorname{deg}(x_j) = (a_j, b_j)$.
Let $I$ be a (not necessarily bihomogeneous) ideal of $S$. Then the ring $R = S \, / \, I$ is not necessarily bigraded, since the ideal can mix monomials of different bidegrees in $S$. However, it is equipped with a bifiltration inherited from the bigrading on $S$:
\begin{equation}
    F^{p,q}\,R = \pi \left( \,\bigoplus_{p' \leq p, q' \leq q} S_{p',q'} \,\right) \,,
\end{equation}
where $\pi(\cdot)$ is the image of $\cdot$ in $R$.

Given such a polynomial ring $R = S \,/\, I$, which defines our geometry as $X=\operatorname{Spec}(R)$, we can construct the (coordinate ring of the) $n$-th jet scheme, $J_n R = \mathbb{C}[J_n X]$, as follows \cite{Nash}. Write $I = (f_1, \cdots, f_r)$.
Then, we expand the $x_j$, and thus the $f_i$, formally as power series in an auxiliary variable $t$:
\begin{equation}\label{eqn:jetexp}
    x_j = \sum_{\alpha = 0}^n x_j^{(\alpha)}t^\alpha \,, \quad f_i \,\,\operatorname{ mod }\,\, t^{n+1} = \sum_{\alpha = 0}^n f_i^{(\alpha)}t^\alpha \,.
\end{equation}
We can consider $S_n = \mathbb{C}[x_1^{(0)}, x_1^{(1)}, \cdots, x_m^{(n)}]$, which has a natural trigrading given by $\operatorname{deg}(x_j^{(\alpha)}) = (\alpha, a_j, b_j)$.
Then, we can define the ideal
\begin{equation}
    I_n = \left(f_1^{(0)}, \cdots, f_1^{(n)}, \cdots, f_r^{(0)}, \cdots, f_r^{(n)}\right)  \,.
\end{equation}
Thus, $J_n R = S_n \,/\, I_n$ is trifiltered, and we can define the, trifiltered, arc space as the direct limit $J_\infty R = \varinjlim_{n} J_n R$, which is the coordinate ring of the arc space $J_\infty X$. See \cite{ArakawaMoreauBook} for more details.

Finally, given a bifiltered ring, there is an associated bigraded ring defined as follows. Let $F^{p,q}\, R$ be a bifiltration on the ring $R$. Then we can define
\begin{equation}
    \operatorname{gr}^{p,q}(R) = \frac{F^{p,q}\,R}{F^{p-1,q}\,R + F^{p,q-1}\,R} \,,
\end{equation}
and then we have a bigraded ring
\begin{equation}
    \operatorname{gr}(R) = \bigoplus_{p,q}\operatorname{gr}^{p,q}(R) \,.
\end{equation}
Similarly, we have the associated trigraded $\operatorname{gr}(J_n(R))$, and there is a Hilbert series carrying fugacities, $p$, $q$, and $T$, for the three gradings:
\begin{equation}
    \operatorname{HS}_{p,q,T}(\operatorname{gr}(J_n R)) \,.
\end{equation}
Upon specializing to $p=q$, we claim that this produces the Macdonald index when $X$ (or $R$) is chosen appropriately.  

Finally, we note that while (the coordinate ring of) the arc space formally consists of an infinite number of generators and relations, we can show that, for any $\ell$, 
\begin{equation}
  \begin{aligned}
    \operatorname{HS}_{q,q,T}&(\operatorname{gr}(J_\infty R)) \\ &\quad= \operatorname{HS}_{q,q,T}(\operatorname{gr}(J_{\ell-d}R))\mod q^{\ell + 1} \,,
  \end{aligned}
\end{equation}

where $d = \operatorname{min}(a_1, \cdots, a_m)$. Therefore, we can determine the Hilbert series to arbitrary order in $q$ by truncating to a jet ring with a finite number of generators and relations.

\section{Proposal: \texorpdfstring{\boldmath{$(A_{k-1}, A_{N-1})$}}{(Ak-1, AN-1)} theory}

We have explained the proposal of \cite{Kang:2025zub} that, for a given 4d $\mathcal{N}=2$ SCFT, there can exist a bifiltered ring $R$, scheme-ifying the Higgs branch, such that the Macdonald index of the 4d theory matches the Hilbert series of the arc space of $R$, as in equation \eqref{eqn:prop}.\footnote{For conciseness, we frequently refer to the right-hand side of \eqref{eqn:prop} as the ``Hilbert series of the arc space'', with the understanding that the bifiltration, associated graded construction and grading conventions are taken as read.} In this section, we provide a geometric condition on a bifiltered polynomial ring $R$ for the scheme $X$ such that it realizes the proposal for the $(A_{k-1}, A_{N-1})$ Argyres--Douglas theories with $\gcd(k,N) = 1$.

Pick a pair of coprime integers $(k \geq 2,N \geq 2)$, and consider the graded polynomial ring 
\begin{equation}
    S = \mathbb{C}[x_2, \cdots, x_k] \,, \quad \operatorname{deg}(x_d) = d \,,
\end{equation}
where the grading structure is fixed by the degree assignment.
We refer to this degree as the $q$-degree. Furthermore, let $P(x_2, \cdots, x_k)$ denote a $q$-degree $N + k + 1$ polynomial. From $P$, we can determine a Jacobian ideal belonging to the ring $S$:
\begin{equation}
    I_P = (\partial_{x_2} P, \cdots, \partial_{x_k} P) \,.
\end{equation}
We want to consider the arc space of the scheme defined by $R = S \,/\, I_P$, where we refer to the additional degree introduced in the jet construction as the $p$-degree.

We are interested, for reasons that will become clear anon, in the existence of a syzygy (relation of relations) of $(q,p)$-degree $(k + N + 2, 1)$. Let $f_{d,p}$ denote the relation that comes from taking the $t^p$ coefficient of $\partial_{x_d} P$ in the jet expansion as in equation \eqref{eqn:jetexp}. The existence of a minimal syzygy of the specified degrees is then equivalent to the existence of a set of non-zero $g_{a,b}$ such that
\begin{equation}\label{eqn:syzygycond}
    \sum_{d = 2}^k \left(g_{d, 1} f_{d,0} + g_{d, 0} f_{d,1} \right) = 0 \,,
\end{equation}
and where the $(q,p)$-degrees are given by
\begin{equation}
    \operatorname{deg}(g_{a,b}) = (k+N+2-a, b ) \,.
\end{equation}
By expanding equation \eqref{eqn:syzygycond} in the monomial basis, it can be written in the form $M g = 0$ where $M$ is called the syzygy matrix.  
We propose that there is a unique ideal $I_P$, for a given $(k, N)$, for which this syzygy condition can be solved (meaning that the rank of $M$ drops by 1), and for which $R = S\,/\,I_P$ is a complete intersection. We refer to such an ideal as $I_P^\text{AD}$. 

Now, we introduce a bigrading on $S$ via the bidegree assignments
\begin{equation}
    \operatorname{deg}(x_d) = (d, d-1) \,.
\end{equation}
The first degree is the $q$-degree, and the second degree we refer to as the $T$-degree. We propose that the Hilbert series of the arc space of $R_{\text{AD}} \equiv S \,/\, I_P^\text{AD}$ is identical to the Macdonald index of the $(A_{k-1}, A_{N-1})$ theory:
\begin{equation}
    \operatorname{HS}_{q,q,T}\left( \operatorname{gr} ( J_\infty R_{\text{AD}})\right) = I_\text{Mac}^{(A_{k-1}, A_{N-1})}(q,T) \,.
\end{equation}
Furthermore, $\operatorname{Spec}(R_{\text{AD}})_\text{red}$ is a point, which is the Higgs branch of such $(A_{k-1}, A_{N-1})$ theories.

We can motivate the conditions on the bifiltered ring $R$ by studying the known Schur index for these Argyres--Douglas theories. The Schur index, $I_\text{S}(q)$, of the $(A_{k-1}, A_{N-1})$ Argyres--Douglas theory, which is the $T \rightarrow 1$ limit of the Macdonald index, is \cite{ Cordova:2015nma, Song:2015wta, Song:2017oew}:
\begin{equation}\label{eqn:SchurGeneral}
    I_S = \operatorname{PE} \left[\frac{q^2 + \cdots + q^k - q^{N + 1} - \cdots - q^{k + N - 1}}{(1-q)(1-q^{k+N})} \right] \,.
\end{equation}
Let us suppose that this Schur index is reproduced via the Hilbert series of the arc space of $R$:
\begin{equation}
    \operatorname{HS}_{q,q}(J_\infty R) = I_\text{S}(q) \,.
\end{equation}
The numerator in the plethystic exponential indicates that we should consider a ring $R$ with $k-1$ generators and relations of degrees, $2, \cdots, k$, and $N + 1, \cdots, k + N - 1$, respectively. The Hilbert series is obtained from the Betti numbers of a minimal free resolution of the arc space $J_\infty R$. By comparing the first terms in the expansion of the Schur index (up to and including order $O(q^{k + N + 3})$), we identify the \textbf{necessary and sufficient} geometric condition required to match the physical theory:
\begin{align}\label{eqn:betti}
\begin{gathered}
    R\text{ is a complete intersection}\,,\\
    \beta_{2, (k+N+1,1)}(J_1 R) = \beta_{2, (k + N + 2, 1)}(J_1 R) = 1 \,.
\end{gathered}
\end{align}
We propose that this is the \emph{defining extremality condition} on the Betti numbers. The Betti number $\beta_{2,(q,p)}$ counts the number of minimal syzygies of bidegree $(q,p)$; the required $p$-degree of the Betti numbers can be extracted from the fact that $R$ is a complete intersection. 
The first requirement ($\beta_{2, (k+N+1,1)} = 1$) is satisfied if and only if $R$ is a complete intersection quotient by a Jacobian ideal. The second requirement ($\beta_{2, (k + N + 2, 1)} = 1$) is satisfied if and only if there exists a unique minimal syzygy as in equation \eqref{eqn:syzygycond}.\footnote{In general, it is also necessary to verify that such an explicit syzygy is minimal and thus that it contributes to the Betti numbers. Fortunately, the graded structure implies that this syzygy, if it exists, cannot be generated by lower-degree syzygies.} Together, these constraints eliminate the continuous deformations of the scheme, uniquely fixing the ideal $I_P$ that maps to the physical theory.

Generally, geometries arising from OPE decouplings possess a moduli space of parameters. It contains various singular loci where the rank of the syzygy matrix drops. We find that there exists a \emph{unique point}, lying at the intersection of these loci, where the Betti numbers are precisely given by equation \eqref{eqn:betti}, while preserving the complete intersection property of the geometry.

\section{Argyres--Douglas examples}

We now use the geometric conditions proposed in the previous section to determine the bifiltered rings that capture the Macdonald index and Higgs branch of a variety of $(A_{k-1}, A_{N-1})$ Argyres--Douglas theories. \\[-8pt]


\noindent\paragraph{\underline{\boldmath{$(A_1, A_{N-1})$} \textbf{theories}}}

The simplest and most straightforward class of theories are the $(A_1, A_{2n})$ theories. We expect only a single generator, $x_2$, of degree two, and a single relation, of degree $2n + 2$. Thus,
\begin{equation}\label{eqn:A1ring}
    R = \mathbb{C}[x_2] \, / \, \left(\, x_2^{n+1} \,\right) \,,
\end{equation}
is the only possibility. 
This can be trivially written as the quotient by the Jacobian ideal of a degree $n + 3$ polynomial. Furthermore, it is easy to see that the conditions on the Betti numbers in equation \eqref{eqn:betti} are automatically satisfied for such an $R$. It has been extensively tested \cite{bai2020quadratic,Andrews:2025krn,Kang:2025zub,Bhargava:2023hsc} that the Hilbert series of the arc space of the ring in equation \eqref{eqn:A1ring} reproduces the Macdonald index of the 4d SCFT.\\[-8pt]


\noindent\paragraph{\underline{\boldmath{$(A_2, A_{N-1})$} \textbf{theories}}}

The $(A_2, A_{N-1})$ theories, where $\gcd(3, N) = 1$, provide the first cases where the geometric conditions in equation \eqref{eqn:betti} impose non-trivial constraints. To illustrate this, we consider the $(A_2, A_9)$ theory in detail. The degrees of the generators and relations, which can be determined from the Schur index, indicate that we should consider a ring
\begin{equation}
    R = \mathbb{C}[x_2, x_3] \, / \, (a x_2^6 + b x_2^3 x_3^2 + c x_3^4, d x_2^4 x_3 + e x_2 x_3^3) \,,
\end{equation}
The condition that there exists a non-trivial syzygy of degree $(14,1)$ in $J_1(R)$ imposes that $be - 8dc = 0$, and, in combination with rescaling and normalization, this condition allows us to write any such solution as a quotient by the Jacobian ideal associated to the polynomial
\begin{equation}
    P(x_2, x_3) = \beta x_2^7 +  x_3^2 x_2^4 + x_3^4 x_2 \,,
\end{equation}
for some free parameter $\beta$. The condition that $\beta_{2,(15,1)} = 1$ further constrains the remaining coefficient to satisfy
\begin{equation}
    21 \beta - 1 = 0 \,.
\end{equation}
In this way, we obtain the ideal $I_P^\text{AD}$ for the $(A_2, A_9)$ Argyres--Douglas theory. Remarkably, very few conditions completely fixed $R$, which precisely reproduces the Schur index as the Hilbert series of the arc space. We have verified that they match up to $O(q^{30})$.

A similar analysis can be carried out for any $(A_2, A_{N-1})$ where $\gcd(3, N) = 1$. We have listed the solutions for $N \leq 23$ in Table \ref{tab:polynomials}. For $N \leq 7$ the expansions of the Macdonald indices have been given in \cite{Song:2017oew}, and we have verified that these expressions match with the Hilbert series of the arc space. In fact, the Hilbert series calculation easily extends to significantly higher-orders than has been written in the literature, and we leave such results as a prediction. We have verified that the Schur index matches the Schur limit of the Hilbert series. We have included some such Hilbert series in the Appendix.\\[-8pt]


\noindent\paragraph{\underline{\boldmath{$(A_3, A_{N-1})$} \textbf{theories}}}

Let us now consider some of the $(A_3, A_{N-1})$ series, with $\gcd(4, N) = 1$. The Jacobian ideals we consider, $I_P$, are obtained from a $q$-degree $N + 5$ polynomial, $P$, belonging to the ring
\begin{equation}
    S = \mathbb{C}[x_2, x_3, x_4] \quad \text{where} \quad \operatorname{deg}(x_d) = d \,.
\end{equation}
The ideal $I_P$ is unchanged under a shift of $P$ via
\begin{equation}\label{eqn:A3shifts}
    x_2 \rightarrow a x_2 \,,\quad x_3 \rightarrow b x_3 \,,\quad x_4 \rightarrow c x_4 + d x_2^2 \,,
\end{equation}

We start with the $(A_3, A_4)$ SCFT. Using the shifts in equation \eqref{eqn:A3shifts}, the polynomial $P$ can be written in terms of a single coefficient, $\beta$:
\begin{align}
   P = \beta x_{2}^{5} + x_{2}^{2} x_{3}^{2} + x_{2} x_{4}^{2} + x_{3}^{2} x_{4} \,.
\end{align}
It is easy to see that a syzygy of $(q,p)$-degree $(11, 1)$ in the ring $J_\infty( S \, / \, I_P )$ only exists if
\begin{equation}
    10\beta - 1 = 0 \,.
\end{equation}
Thus, we have determined the ideal $I_P^\text{AD}$, and we can compare the Hilbert series of the arc space of $R = S \, / \, I_P^\text{AD}$ and observe it matches the indices of the $(A_3, A_4)$ SCFT.

Next, we can consider the $(A_3, A_6)$ theory. We can use the shifts in equation \eqref{eqn:A3shifts} to write the polynomial as 
\begin{align}
    \beta_1 x_{2}^{6} + \beta_2 x_{2}^{4} x_{4} + \beta_3 x_{3}^{4} + x_{2}^{3} x_{3}^{2} + x_{2} x_{3}^{2} x_{4} + x_{4}^{3} \,.
\end{align}
It turns out that there are multiple branches of solutions in which the syzygy exists. However, we find that there exists a unique solution (up to shifts and reparameterization), where $R$ is a complete intersection to be
\begin{align}
  \beta_1 = 0 \,, \quad \beta_2 = \frac{9}{2} \,, \quad \beta_3 = \frac{1}{108}  \,.
\end{align}
These conditions define the ideal $I_P^\text{AD}$, such that the Schur limit of the Hilbert series of the arc space of $S \, / \, I_P^\text{AD}$ reproduces the Schur index of the $(A_3, A_6)$ theory.

\begin{table}[tbp]
\centering
\small
\begin{tabular}{cl}
\toprule
Theory & Polynomial $P(x_2, x_3, \cdots)$ \\
\midrule
$(A_2, A_3)$ & $x_{2}^{4} + x_{2} x_{3}^{2}$  \\
$(A_2, A_4)$ & $x_{2}^{3} x_{3} + x_{3}^{3}$  \\
$(A_2, A_6)$ & $x_{2}^{4} x_{3} + x_{2} x_{3}^{3}$  \\
$(A_2, A_7)$ & $\frac{1}{30}x_{2}^{6} + x_{2}^{3} x_{3}^{2} + x_{3}^{4}$  \\
$(A_2, A_9)$ & $\frac{1}{21}x_{2}^{7} + x_{2}^{4} x_{3}^{2} + x_{2} x_{3}^{4}$  \\
$(A_2, A_{10})$ & $x_{2}^{6} x_{3} + x_{2}^{3} x_{3}^{3} + \frac{3}{50} x_{3}^{5}$  \\
$(A_2, A_{12})$ & $\frac{3}{35} x_{2}^{7} x_{3} + x_{2}^{4} x_{3}^{3} + x_{2} x_{3}^{5}$  \\
$(A_2, A_{13})$ & $\frac{5}{72} x_{2}^{9} + x_{2}^{6} x_{3}^{2} + x_{2}^{3} x_{3}^{4} + \frac{2}{25} x_{3}^{6}$  \\
$(A_2, A_{15})$ & $\frac{8}{7875} x_{2}^{10} + \frac{4}{35} x_{2}^{7} x_{3}^{2} + x_{2}^{4} x_{3}^{4} + x_{2} x_{3}^{6}$  \\
$(A_2, A_{16})$ & $\frac{1}{8} x_{2}^{9} x_{3} + x_{2}^{6} x_{3}^{3} + x_{2}^{3} x_{3}^{5} + \frac{2}{21} x_{3}^{7}$  \\
$(A_2, A_{18})$ & $\frac{8}{3087} x_{2}^{10} x_{3} + \frac{20}{147} x_{2}^{7} x_{3}^{3} + x_{2}^{4} x_{3}^{5} + x_{2} x_{3}^{7}$  \\
$(A_2, A_{19})$ & $\frac{1}{396} x_{2}^{12} + \frac{1}{6} x_{2}^{9} x_{3}^{2} + x_{2}^{6} x_{3}^{4} + x_{2}^{3} x_{3}^{6} + \frac{3}{28} x_{3}^{8}$  \\
$(A_2, A_{21})$ & $\frac{3}{249704} x_{2}^{13} + \frac{3}{686} x_{2}^{10} x_{3}^{2} + \frac{15}{98} x_{2}^{7} x_{3}^{4} + x_{2}^{4} x_{3}^{6} + x_{2} x_{3}^{8}$  \\
$(A_2, A_{22})$ & $\frac{125}{19404} x_{2}^{12} x_{3} + \frac{25}{126} x_{2}^{9} x_{3}^{3} + x_{2}^{6} x_{3}^{5} + x_{2}^{3} x_{3}^{7} + \frac{7}{60} x_{3}^{9}$  \\
\midrule
$(A_3, A_4)$ & $\frac{1}{10}x_{2}^{5} + x_{2}^{2} x_{3}^{2} + x_{2} x_{4}^{2} + x_{3}^{2} x_{4}$  \\
$(A_3, A_6)$ & $\frac{9}{2} x_2^4 x_4 + x_2^3 x_3^2 + x_2 x_3^2 x_4 + \frac{1}{108} x_3^4 + x_4^3$  \\
$(A_3, A_8)$ & $\frac{x_{2}^{5} x_{4}}{120} + \frac{x_{2}^{4} x_{3}^{2}}{72} + \frac{x_{2}^{2} x_{3}^{2} x_{4}}{3} + \frac{x_{2} x_{3}^{4}}{18} + x_{2} x_{4}^{3} + x_{3}^{2} x_{4}^{2}$ \\
$(A_3, A_{10})$ & $\frac{x_{2}^{8}}{47029248} + \frac{x_{2}^{5} x_{3}^{2}}{7776} + \frac{x_{2}^{4} x_{4}^{2}}{1296} + \frac{x_{2}^{3} x_{3}^{2} x_{4}}{54} + \frac{x_{2}^{2} x_{3}^{4}}{18} $ \\ 
& $\quad + x_{2} x_{3}^{2} x_{4}^{2} + x_{3}^{4} x_{4} + x_{4}^{4}$ \\
\midrule
$(A_4, A_5)$ & $\frac{61776859}{5} x_{2}^{6} + 184899 x_{2}^{4} x_{4} + 10368 x_{2}^{3} x_{3}^{2} + x_{2}^{2} x_{3} x_{5} $ \\
& ~~ $+ 861 x_{2}^{2} x_{4}^{2} + x_{2} x_{3}^{2} x_{4} + x_{2} x_{5}^{2} + x_{3}^{4} + x_{3} x_{4} x_{5} + x_{4}^{3}$ \\
\bottomrule
\end{tabular}
\caption{Polynomials for $(A_{k-1}, A_{N-1})$ theories. $P(x_2,x_3,...)$ is a homogeneous polynomial of the $x_d$, with degree $k+N+1$, where $x_d$ is of degree $d$. The coefficients are fixed by demanding $\beta_{2,(k+N+2,1)}(J_1R)=1$.}\label{tab:polynomials}
\end{table}

It becomes more expensive to perform computation for the higher-rank case, since it requires computing the determinantal variety of a large syzygy matrix. We summarize some explicit results in Table \ref{tab:polynomials}.

\section{\texorpdfstring{Vertex Operator Algebras \\ and Null Relations}{Vertex Operator Algebras and Null Relations}}

As mentioned in the introduction, the bifiltered rings that capture the Higgs branch and Macdonald index of a 4d $\mathcal{N}=2$ SCFT are expected to be derived from decoupling phenomena in the OPEs of quarter-BPS operators. Under the SCFT/VOA correspondence \cite{Beem:2013sza}, such a decoupling relation corresponds to a null relation in the associated VOA. Therefore, the geometric conditions on the rings that we proposed in equation \eqref{eqn:betti} should have an interpretation in terms of null relations in vertex operator algebras. We discuss this connection in this section.

For many of the Argyres--Douglas theories we consider in this letter, the VOAs are (partially) known, and thus we can, in principle, understand the bifiltered ring.\footnote{In fact, in \cite{Xie:2019zlb}, it was conjectured that for such $(A_{k-1}, A_{N-1})$ theories, Zhu's $C_2$-algebra, a certain truncation of the VOA, is given by a Jacobi algebra defined by a polynomial of fixed degree. However, the coefficients were not given. 
}

For example, the OPE of $(n+1)$ copies of the stress-tensor supermultiplet, $\widehat{\mathcal{C}}_{0(0,0)}$, generically, by symmetry principles, contains a $\widehat{\mathcal{C}}_{n(\frac{n}{2}, \frac{n}{2})}$ supermultiplet.\footnote{We use the notation of \cite{Dolan:2002zh} to denote the supermultiplets of the 4d superconformal algebras.} For the $(A_1, A_{2n})$ Argyres--Douglas theory, this multiplet is absent from the OPE \cite{Agarwal:2018zqi,Liendo:2015ofa}. In the language of the associated VOA, which is the Virasoro minimal model $M(2, 2n + 3)$, the $\widehat{\mathcal{C}}_{0(0,0)}$ multiplet maps to the $L_{-2}$ strong generator, and the absence of the $\widehat{\mathcal{C}}_{n(\frac{n}{2}, \frac{n}{2})}$ multiplet from the OPE is related to a null relation:
\begin{equation}
    ((L_{-2})^{n+1} + \cdots ) \big| \Omega, c \big\rangle \,,
\end{equation}
where $\Omega$ is the vacuum state, and $c$ is the central charge. The connection between this null relation and the ring in equation \eqref{eqn:A1ring} is clear.

For $k > 2$ the situation becomes more complicated. The explicit expressions for the null vectors in the W-algebra $W(k, k + N)$, which is the VOA for $(A_{k-1}, A_{N-1})$ when $\gcd(k,N) = 1$ \cite{Buican:2015ina,Cordova:2015nma}, are generally not known. 

For $k = 3$, leading null relations, for certain values of the central charge, have been determined in \cite{Agarwal:2018zqi}. In these cases there are two strong generators $L_{-2}$ and $W_{-3}$. For $c = -114/7$, corresponding to the $(A_2, A_3)$ theory, the null relations are \cite{Agarwal:2018zqi}:
\begin{equation}
    \begin{aligned}
        \left( W_{-3} L_{-2} + \cdots \right) &\big| \Omega, c \big\rangle \,, \\
        \left( (L_{-2})^3 - \frac{39}{7}(W_{-3})^2  + \cdots \right) &\big| \Omega, c \big\rangle \,,
    \end{aligned}
\end{equation}
where $| \Omega, c \rangle$ is the vacuum state and the $\cdots$ contain terms involving descendants. Under the mapping $L_{-2} \rightarrow x_2$ and $W_{-3} \rightarrow x_3$, we recover the ideal given in Table \ref{tab:polynomials}.

For $N = 5, 7$ the ideals in Table \ref{tab:polynomials} are also recovered from the null relations of the $W(3, 3 + N)$ algebra, as presented in \cite{Agarwal:2018zqi}. For $N > 7$ the relevant null relations have not been determined; thus we conjecture that the ideals in Table \ref{tab:polynomials} reproduce those null relations. 

More generally, we propose that the $W(k, k+N)$ vertex operator algebra (which has strong generators $W_{-2}$, $W_{-3}$, $\cdots$, $W_{-k}$) 
has null relations of the form
\begin{equation}\label{eqn:nullpred}
  \begin{aligned}
    \left( \partial_{x_d} P(x_2, \cdots, x_k)|_{x_j\rightarrow W_{-j}} + \cdots \right) \big| \Omega, c \big\rangle \,,
  \end{aligned}
\end{equation}
for $d = 2, \cdots, k$. Here, $P$ is a polynomial of $q$-degree $k + N + 1$ such that $\mathbb{C}[x_2, \cdots, x_k] \, / \, I_P$ satisfies equation \eqref{eqn:betti}. That is, the numerical coefficients appearing in the null relations are determined via the simple geometric condition in equation \eqref{eqn:betti}. We leave this as a general prediction of our work.

\section{Discussion}

We have presented evidence that, for a specific class of 4d $\mathcal{N}=2$ SCFTs, the Macdonald index and Higgs branch are completely determined by a bifiltered affine scheme subject to an extremality condition, which manifests as the existence of specific syzygies in the associated jet scheme. This result suggests a new geometric paradigm for identifying physical theories. Two natural questions arise regarding the scope of this correspondence:
\begin{itemize}
	\item Does the bifiltered affine scheme determine the full vertex operator algebra, and thus the complete conformal data of the quarter-BPS sector?
	\item Does this scheme uniquely characterize the entire SCFT? That is, modulo marginal deformations and global structure, is the 4d theory fully fixed by this geometric data?\footnote{There is no known example of two distinct $\mathcal{N}=2$ SCFTs (up to marginal deformations and global structure) giving rise to the same associated VOA.}
\end{itemize}

In this letter, we have focused on 4d $\mathcal{N}=2$ SCFTs where the Higgs branch is simply a point. However, bifiltered affine schemes satisfying the proposal in equation~\eqref{eqn:prop} have also been identified for Argyres--Douglas theories with non-trivial Higgs branches \cite{Kang:2025zub, Andrews:2025krn, KLStoAppear}. 
A crucial open problem is to determine if the extremality principle discovered here is universal. We conjecture that these more general schemes are also determined by an explicit extremization principle: specifically, 
that physical theories with non-trivial Higgs branches correspond to unique singular points in the moduli space of schemes where the topology of the jet scheme undergoes a discontinuous change. Verifying this would establish ``geometric extremality'' as a robust selection rule for the landscape of $\mathcal{N}=2$ SCFTs.

\vspace{0.3cm}
\begin{acknowledgments}
We thank Tomoyuki Arakawa, Pieter Bomans, Jacques Distler, Grant Elliot, Heeyeon Kim, and Lorenzo Mansi for useful discussions. 
C.L.~and J.S.~thank Texas A\&M for hospitality. M.J.K.~is supported by the Start-up Research Grant for new faculty provided by Texas A\&M University. C.L.~acknowledges support from DESY (Hamburg, Germany), a member of the Helmholtz Association HGF; C.L.~also acknowledges the Deutsche Forschungsgemeinschaft under Germany's Excellence Strategy - EXC 2121 ``Quantum Universe'' - 390833306 and the Collaborative Research Center - SFB 1624 ``Higher Structures, Moduli Spaces, and Integrability'' - 506632645. The work of J.S.~is supported by the National Research Foundation of Korea (NRF) grants RS-2023-00208602 and RS-2024-00405629, and also by the KAIST-KIAS collaboration program and the Walter Burke Institute of Theoretical Physics at Caltech, and by the U.S. Department of Energy, under Award Number DE-SC0011632.
\end{acknowledgments}

\bibliography{references}
\newpage
\onecolumngrid
\begin{appendix}

\section{Appendix: Macdonald Index as a Hilbert Series}\label{app:HS}

In this appendix, we collect the expressions for the expansions of various Hilbert series on the right-hand side of equation \eqref{eqn:prop}. We list the Hilbert series up to the order determined by {\tt Macaulay2} using approximately one minute of single-core runtime on a modern laptop.\footnote{Due to the double-exponential complexity involved in determining a Gr\"obner basis \cite{MR683204}, the time required to compute the next order in the Hilbert series grows quickly. However, the algorithm we use is not optimized, so there is substantial room for speedup.}
Here, we include the Hilbert series relevant for the $(A_2, A_{N-1})$ theories, for $N = 4, 5, 7, 8, 10, 11$. In each case, the bifiltered rings were given in Table \ref{tab:polynomials}.

We start with the $(A_2, A_3)$ theory, for which the Hilbert series of the arc space is:
\begin{equation}\label{eqn:A2A3HS}
  \begin{aligned}
    &\operatorname{HS}_{q,q,T}(\operatorname{gr}(J_\infty R)) = 1 + q^2 T + q^3(T^2 + T) + q^4(2T^2 + T) + q^5(2T^2 + T) + q^6(2T^3 + 3T^2 + T) \\ &\quad + q^7(3T^3 + 3T^2 + T) + q^8(T^4 + 5T^3 + 4T^2 + T) + q^9(2T^4 + 7T^3 + 4T^2 + T) \\ &\quad+ q^{10}(5T^4 + 9T^3 + 5T^2 + T) + q^{11}(8T^4 + 11T^3 + 5T^2 + T) \\ &\quad+ q^{12}(2T^5 + 13T^4 + 14T^3 + 6T^2 + T) + q^{13}(4T^5 + 17T^4 + 16T^3 + 6T^2 + T) + \cdots \,.
  \end{aligned}
\end{equation}
This calculation, up to and including the $q^{13}$ term, can be determined by truncating to $J_{11} R$. The Macdonald index of the $(A_2, A_3)$ has been determined from the perspective of the $W_3$-algebra in \cite{Agarwal:2018zqi,Foda:2019guo}. See also \cite{Kim:2025klh}. The closed-form expression is
\begin{equation}\label{eqn:A2A3Mac}
    I_\text{Mac}(q, T) = \sum_{\substack{n_1, n_2, n_3, n_4 \geq 0}} \frac{q^{(n_1+n_2+n_3)^2+(n_2 +n_3)^2+n_3^2+n_4^2+(n_1 +2n_2+3n_3 )n_4+n_1 +2n_2+3n_3 +2n_4}}{(q)_{n_1}(q)_{n_2}(q)_{n_3}(q)_{n_4}} T^{n_1 +2n_2+3n_3 +2n_4} \,.
\end{equation}
We can see that the power-series expansion of this expression matches the Hilbert series given in equation \eqref{eqn:A2A3HS}. We leave it as an open question to prove that the closed-form expression for the Hilbert series of the arc space of the ring in Table \ref{tab:polynomials} is as in equation \eqref{eqn:A2A3Mac}.

For the $(A_2, A_4)$ theory, we consider the truncation to $J_{11}R$ to find the expansion:
\begin{equation}\label{eqn:A2A4HS}
  \begin{aligned}
    &\operatorname{HS}_{q,q,T}(\operatorname{gr}(J_\infty R)) = 1 + q^2 T 
    + q^3(T^2 + T)  
    + q^4(2T^2 + T) 
    + q^5(T^3 + 2T^2 + T) 
    + q^6(3T^3 + 3T^2 + T) \\ &\quad
    + q^7(4T^3 + 3T^2 + T) 
    + q^8(3T^4 + 6T^3 + 4T^2 + T) 
    + q^9(5T^4 + 8T^3 + 4T^2 + T) \\ &\quad
    + q^{10}(T^5 + 9T^4 + 10T^3 + 5T^2 + T)
    + q^{11}(3T^5 + 13T^4 + 12T^3 + 5T^2 + T) \\ &\quad
    + q^{12}(8T^5 + 19T^4 + 15T^3 + 6T^2 + T) 
    + q^{13}(14T^5 + 24T^4 + 17T^3 + 6T^2 + T) + \cdots \,.
  \end{aligned}
\end{equation}
This matches with the expression for the refined character of the $W_3$ algebra at $c = -23$, as given up to and including the $q^9$ term in \cite{Agarwal:2018zqi}.

For $(A_2, A_6)$, we compute using $J_{13}R$, and we find the Hilbert series has the following expansion:
\begin{equation}\label{eqn:A2A6HS}
  \begin{aligned}
    &\operatorname{HS}_{q,q,T}(\operatorname{gr}(J_\infty R)) = 1 + q^2 T 
    + q^3(T^2 + T)  
    + q^4(2T^2 + T) 
    + q^5(T^3 + 2T^2 + T) 
    + q^6(T^4 + 3T^3 + 3T^2 + T)  \\ &\quad
    + q^7(2T^4 + 4T^3 + 3T^2 + T)
    + q^8(5T^4 + 6T^3 + 4T^2 + T)
    + q^9(2T^5 + 7T^4 + 8T^3 + 4T^2 + T) \\ &\quad
    + q^{10}(6T^5 + 11T^4 + 10T^3 + 5T^2 + T) 
    + q^{11}(T^6 + 10T^5 + 15T^4 + 12T^3 + 5T^2 + T) \\ &\quad
    + q^{12}(6T^6 + 17T^5 + 21T^4 + 15T^3 + 6T^2 + T) 
    + q^{13}(11T^6 + 25T^5 + 26T^4 + 17T^3 + 6T^2 + T) \\ &\quad
    + q^{14}(2T^7 + 21T^6 +  36T^5 + 34T^4 + 20T^3 + 7T^2 + T) \\ &\quad
    + q^{15}(7T^7 + 34T^6 +  49T^5 + 41T^4 + 23T^3 + 7T^2 + T) + \cdots \,.
  \end{aligned}
\end{equation}
Again, this matches with the expression for the refined character of the $W_3$ algebra at $c = -186/5$, which was given up to and including the $q^9$ term in \cite{Agarwal:2018zqi}.

Next, we turn to the $(A_2, A_7)$ SCFT. In this case, we use $J_{14}R$ and we find the following Hilbert series expansion:
\begin{equation}\label{eqn:A2A7HS}
  \begin{aligned}
    &\operatorname{HS}_{q,q,T}(\operatorname{gr}(J_\infty R)) = 1 + q^2 T 
    + q^3(T^2 + T)  
    + q^4(2T^2 + T) 
    + q^5(T^3 + 2T^2 + T) 
    + q^6(T^4 + 3T^3 + 3T^2 + T) \\ &\quad
    + q^7(2T^4 + 4T^3 + 3T^2 + T) 
    + q^8(T^5 + 5T^4 + 6T^3 + 4T^2 + T) 
    + q^9(3T^5 + 7T^4 + 8T^3 + 4T^2 + T) \\ &\quad
    + q^{10}(7T^5 + 11T^4 + 10T^3 + 5T^2 + T) 
    + q^{11}(3T^6 + 11T^5 + 15T^4 + 12T^3 + 5T^2 + T) \\ &\quad
    + q^{12}(9T^6 + 18T^5 + 21T^4 + 15T^3 + 6T^2 + T) 
    + q^{13}(T^7 + 15T^6 + 26T^5 + 26T^4 + 17T^3 + 6T^2 + T) \\ &\quad
    + q^{14}(7T^7 + 26T^6 +  37T^5 + 34T^4 + 20T^3 + 7T^2 + T) \\ &\quad
    + q^{15}(15T^7 + 40T^6 +  50T^5 + 41T^4 + 23T^3 + 7T^2 + T) \\ &\quad
    + q^{16}(3T^8 + 29T^7 + 59T^6 +  66T^5 + 51T^4 + 26T^3 + 8T^2 + T) + \cdots \,.
  \end{aligned}
\end{equation}
The Macdonald index in \cite{Agarwal:2018zqi}, which is given up to and including the $q^9$ term, matches with this Hilbert series.

Finally, we turn to the $(A_2, A_9)$ and $(A_2, A_{10})$ theories. For these theories, the Macdonald index has not been written down explicitly, as the null relations in the associated vertex operator algebras are challenging to work out. We have compared the $T \rightarrow 1$ limit of the Hilbert series to the Schur indices in these cases. For $(A_2, A_9)$, the Hilbert series of the arc space is
\begin{equation}\label{eqn:A2A9HS}
  \begin{aligned}
    &\operatorname{HS}_{q,q,T}(\operatorname{gr}(J_\infty(R))) = 1 + q^2 T 
    + q^3(T^2 + T) 
    + q^4(2T^2 + T) 
    + q^5(T^3 + 2T^2 + T) 
    + q^6(T^4 + 3T^3 + 3T^2 + T) \\ &\quad
    + q^7(2T^4 + 4T^3 + 3T^2 + T) 
    + q^8(T^5 + 5T^4 + 6T^3 + 4T^2 + T) 
    + q^9(T^6 + 3T^5 + 7T^4 + 8T^3 + 4T^2 + T) \\ &\quad
    + q^{10}(2T^6 + 7T^5 + 11T^4 + 10T^3 + 5T^2 + T) 
    + q^{11}(5T^6 + 11T^5 + 15T^4 + 12T^3 + 5T^2 + T) \\ &\quad
    + q^{12}(2T^7 + 11T^6 + 18T^5 + 21T^4 + 15T^3 + 6T^2 + T) 
    + q^{13}(6T^7 + 17T^6 + 26T^5 + 26T^4 + 17T^3 + 6T^2 + T) \\ &\quad
    + q^{14}(T^8 + 14T^7 + 28T^6 +  37T^5 + 34T^4 + 20T^3 + 7T^2 + T) \\ &\quad
    + q^{15}(6T^8 + 24T^7 + 42T^6 +  50T^5 + 41T^4  + 23T^3 + 7T^2 + T) \\ &\quad
    + q^{16}(16T^8 + 40T^7 + 61T^6 +  66T^5  + 51T^4 + 26T^3 + 8T^2 + T) \\ &\quad
    + q^{17}(2T^9 + 29T^8 + 62T^7 + 84T^6 +  84T^5  + 60T^4 + 29T^3 + 8T^2 + T) \\ &\quad
    + q^{18}(12T^9 + 52T^8 + 93T^7 + 116T^6 +  107T^5  + 72T^4 + 33T^3 + 9T^2 + T) + \cdots \,.
  \end{aligned}
\end{equation}
Since the Schur index of $(A_2, A_9)$ was not studied from the perspective of the Hilbert series of the arc space in \cite{Kang:2025zub}, we write the first coefficients of the Schur limit of the Hilbert series here. They are
\begin{equation}
  \begin{aligned}
    &(1, 0, 1, 2, 3, 4, 8, 10, 17, 24, 36, 49, 74, 99, 142, 194, 269, 359, 495, 654)\,.
  \end{aligned}
\end{equation}
As expected from our discussion, these coefficients agree with the closed-form expression for the Schur index of $(A_2, A_9)$, up to and including the $q^{19}$ term.

Finally, for $(A_2, A_{10})$, the relevant Hilbert series of the arc space is
\begin{equation}\label{eqn:A2A10HS}
  \begin{aligned}
    &\operatorname{HS}_{q,q,T}(\operatorname{gr}(J_\infty(R))) = 1 + q^2 T 
    + q^3(T^2 + T)  
    + q^4(2T^2 + T) 
    + q^5(T^3 + 2T^2 + T) 
    + q^6(T^4 + 3T^3 + 3T^2 + T) \\ &\quad
    + q^7(2T^4 + 4T^3 + 3T^2 + T) 
    + q^8(T^5 + 5T^4 + 6T^3 + 4T^2 + T) 
    + q^9(T^6 + 3T^5 + 7T^4 + 8T^3 + 4T^2 + T) \\ &\quad
    + q^{10}(2T^6 + 7T^5 + 11T^4 + 10T^3 + 5T^2 + T) 
    + q^{11}(T^7 + 5T^6 + 11T^5 + 15T^4 + 12T^3 + 5T^2 + T) \\ &\quad
    + q^{12}(T^8 + 2T^7 + 11T^6 + 18T^5 + 21T^4 + 15T^3 + 6T^2 + T) \\ &\quad
    + q^{13}(T^8 + 6T^7 + 17T^6 + 26T^5 + 26T^4 + 17T^3 + 6T^2 + T) \\ &\quad
    + q^{14}(T^9 + 3T^8 + 14T^7 + 28T^6 +  37T^5 + 34T^4  + 20T^3 + 7T^2 + T) \\ &\quad
    + q^{15}(T^{10} + 2T^9 + 7T^8 + 24T^7 + 42T^6 +  50T^5 + 41T^4 + 23T^3 + 7T^2 + T) \\ &\quad
    + q^{16}(T^{10} + 4T^9 + 17T^8 + 40T^7 + 61T^6 +  66T^5  + 51T^4 + 26T^3 + 8T^2 + T) \\ &\quad
    + q^{17}(T^{11} + 2T^{10} + 8T^9 + 30T^8 + 62T^7 + 84T^6 +  84T^5  + 60T^4 + 29T^3 + 8T^2 + T) \\ &\quad
    + q^{18}(T^{12} + 2T^{11} + 5T^{10} + 17T^9 + 53T^8 + 93T^7  + 116T^6 +  107T^5 + 72T^4  + 33T^3 + 9T^2 + T) \\ &\quad
    + q^{19}(T^{12} + 4T^{11} + 10T^{10} + 32T^9 + 84T^8  + 132T^7  + 151T^6 +  132T^5  + 83T^4 + 36T^3 + 9T^2 + T) + \cdots \,,
  \end{aligned}
\end{equation}
which we propose to be identical to the leading orders of the Macdonald index of the $(A_2, A_{10})$ Argyres--Douglas SCFT. We also list the first coefficients (up to and including $q^{20}$) in the Schur limit, $T \rightarrow 1$, of the Hilbert series:
\begin{equation}
  \begin{aligned}
    &(1, 0, 1, 2, 3, 4, 8, 10, 17, 24, 36, 50, 75, 100, 145, 198, 275, 370, 510, 676, 916) \,.
  \end{aligned}
\end{equation}
This matches with the expansion of the known closed-form expression for the $(A_2, A_{10})$ Schur index.

\end{appendix}

\end{document}